\documentclass[10pt,journal,compsoc]{IEEEtran}
\usepackage{CJK}

\ifCLASSOPTIONcompsoc
  \usepackage[nocompress]{cite}
\else
  \usepackage{cite}
\fi

\ifCLASSINFOpdf

\else

\fi

\usepackage{graphicx}
\usepackage{epstopdf}
\usepackage{booktabs}
\usepackage{subfigure}
\usepackage{amsfonts}
\usepackage{amsmath}
\usepackage{algorithm, algorithmic}
\usepackage{subfigure}
\usepackage{cite}
\usepackage{ragged2e}

\usepackage{amsthm}
\usepackage[T1]{fontenc}
\usepackage[utf8]{inputenc}
\usepackage{authblk}
\usepackage{float}
\usepackage{hyperref}

\hypersetup{hypertex=true,
            colorlinks=true,
            linkcolor=blue,
            anchorcolor=blue,
            citecolor=blue}
\begin{document}
%\begin{CJK*}{GB}{}
\title{Anonymous Pattern Molecular Fingerprint and its Applications on Property Identification}
\author[1, 3, \dag]{Xue Liu}
\author[2, 3, \dag]{Qian Cheng \thanks{\dag These authors contributed to the work equally and should be regarded as co-first authors}}
\author[2, 3]{Dan Sun}
\author[2, 3]{Xing Li}
\author[2, 1, 3, 4, *]{Wei Wei \thanks{* Corresponding author: weiw@buaa.edu.cn}}
\author[1, 3, 4]{Zhiming Zheng}

\affil[1]{Institute of Artificial Intelligence, Beihang University, Beijing, 100191, P. R. China}
\affil[2]{School of Mathematical Sciences, Beihang University, Beijing, 100191, P. R. China}
\affil[3]{Key Laboratory of Mathematics, Informatics and Behavioral Semantics, Ministry of Education, 100191, P. R. China}
\affil[4]{Zhongguancun Laboratory, Beijing, 100094, P. R. China}

% The paper headers
\markboth{PREPRINT SUBMITTED TO arXiv}%
{Shell \MakeLowercase{\textit{et al.}}: Bare Demo of IEEEtran.cls for Computer Society Journals}

\IEEEtitleabstractindextext{%
\begin{abstract}
\justifying Molecular fingerprints are significant cheminformatics tools to map molecules into vectorial space according to their characteristics in diverse functional groups, atom sequences, and other topological structures. In this paper, we set out to investigate a novel molecular fingerprint \emph{Anonymous-FP} that possesses abundant perception about the underlying interactions shaped in small, medium, and large molecular scale links. In detail, the possible inherent atom chains are sampled from each molecule and are extended in a certain anonymous pattern. After that, the molecular fingerprint \emph{Anonymous-FP} is encoded in virtue of the Natural Language Processing technique \emph{PV-DBOW}. \emph{Anonymous-FP} is studied on molecular property identification and has shown valuable advantages such as rich information content, high experimental performance, and full structural significance. During the experimental verification, the scale of the atom chain or its anonymous manner matters significantly to the overall representation ability of \emph{Anonymous-FP}. Generally, the typical scale $r = 8$ enhances the performance on a series of real-world molecules, and specifically, the accuracy could level up to above $93\%$ on all NCI datasets.
\end{abstract}

% Note that keywords are not normally used for peerreview papers.
\begin{IEEEkeywords}
molecular fingerprint, random walks, anonymous pattern walks, typical scale, property identification
\end{IEEEkeywords}}

% make the title area
\maketitle
%\end{CJK*}

\IEEEraisesectionheading{\section{Introduction}\label{sec:introduction}}

\begin{figure}[t]
\centering
\includegraphics[height=7.75cm,width=9cm]{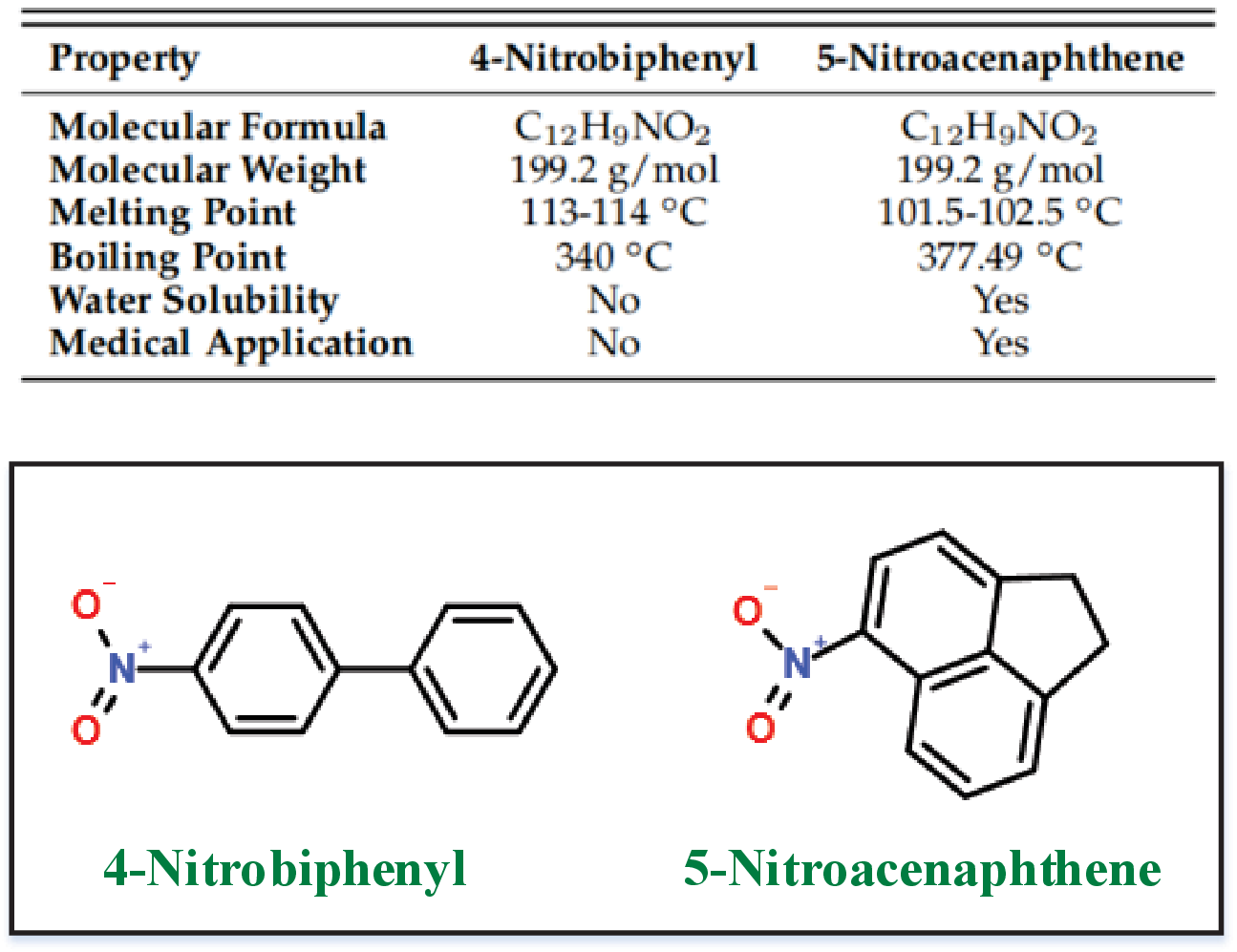}
\caption{The structural graphs for isomers 4-Nitrobiphenyl and 5-Nitroacenaphthene molecules. 4-Nitrobiphenyl is one kind of aromatic compound and 5-Nitroacenaphthene belongs to heteroaromatic compounds.
}
\label{Figure-Isomery}
\end{figure}

The discovery of new molecules benefits human society greatly, and accurate prediction for unknown molecular properties remains an open challenge. In pharmaceutical chemistry~\cite{pharmaceutical-chemistry}, drug designs~\cite{drugDesign}, bioinformatics~\cite{Graph-classification-using-structural-attention} et al., molecules from all of these domains and many more could be represented as graphs, in which nodes interact with others according to the edges among them and integrate as a whole to perform specific properties. In the molecular planar graph, the positions of nodes are occupied by different atoms, and the edges are formed according to the chemical bonds between atoms. The differences in topological structure lead to diversity in chemical or physical properties. For instance, Figure~\ref{Figure-Isomery} introduces the typical isomerism~\cite{isomerism} between the 4-Nitrobiphenyl and 5-Nitroacenaphthene molecules, which are involved in MUTAG dataset~\cite{dataset-mutag} and are labeled differently. Both of them possess a same molecular formula $\textrm{C}_{12}\textrm{H}_{9}\textrm{NO}_{2}$ but express quite different properties (shown in table) due to distinct carbon chain structures. In particular, 5-Nitroacenaphthene is wildly used in pharmaceutical engineering as an important raw material for medicine synthesis.

How to infer the physicochemical properties of molecules, or to distinguish different molecules only from graph topology has recently received a lot of attention from various fields of machine learning. And the core of all these inevitably attributes to the graph isomorphism problem (abbreviated as \emph{\textbf{GIP}})~\cite{Graph-Isomorphism, Graph-Isomorphism-2}. A graph is isomorphic to another if there exists a bijective mapping $f$ of the vertices in this graph to vertices of the other one such that the adjacency could be preserved, i.e., for graph $G_{1} = (\mathcal{V}_{1}, \mathcal{E}_{1})$ and $G_{2} = (\mathcal{V}_{2}, \mathcal{E}_{2})$, for all $v_{i}, v_{j} \in \mathcal{V}_{1}$, $i \neq j$,
\begin{equation}\label{graph-isomorphism}
    (v_{i}, v_{j}) \in \mathcal{E}_{1} \Leftrightarrow (f(v_{i}), f(v_{j})) \in \mathcal{E}_{2}.
\end{equation}

%\begin{table}[t]
%\centering
%\caption{Physical properties and medical applications for isomers 4-Nitrobiphenyl and 5-Nitroacenaphthene.}
%{
%\begin{tabular}{lcc}
%\toprule
%\toprule
%\textbf{Property}                   & \textbf{4-Nitrobiphenyl}                          & \textbf{5-Nitroacenaphthene}                  \\
%\midrule
%\textbf{Molecular Formula}          & $\textrm{C}_{12}\textrm{H}_{9}\textrm{NO}_{2}$    & $\textrm{C}_{12}\textrm{H}_{9}\textrm{NO}_{2}$\\
%\textbf{Molecular Weight}           & 199.2 g/mol                                       & 199.2 g/mol                                   \\
%\textbf{Melting Point}              & 113-114 $^{\circ}$C                               & 101.5-102.5 $^{\circ}$C                       \\
%\textbf{Boiling Point}              & 340 $^{\circ}$C                                   & 377.49 $^{\circ}$C                            \\
%\textbf{Water Solubility}           & No                                                & Yes                                           \\
%\textbf{Medical Application}        & No                                                & Yes                                           \\
%\bottomrule
%\end{tabular}}
%\label{Table-Isomery}
%\end{table}

%\begin{figure}[htbp]\label{Figure-Petersen}
%\centering
%\includegraphics[height=2.5cm,width=7cm]{Figure-Petersen.eps}
%\caption{
%An example of graph isomorphism. (a) represents the Petersen graph~\cite{Petersen-graph}, which is a connected graph with $10$ nodes and $15$ edges. (b) and (c) are both isomorphic to Petersen graph (a) since these three adjacent matrices of (a), (b) and (c) are pairwise similar (or equivalent).
%}
%\end{figure}

Clearly, graph isomorphism problem (\emph{GIP}) is in the class of NP, and subgraph isomorphism has been proved as NP-complete, while it is still unknown whether graph isomorphism belongs to NP-complete or not~\cite{NP-completeness}. Thus it is not feasible to directly apply theoretical fruits of \emph{GIP} into real molecular similarity measure. In recent years, abundant literature in molecular fingerprints have provided effective ways for molecule representation, identification, and comparison, and Maccs~\cite{Fingerprint-MACCS}, PubChem fingerprints~\cite{Fingerprint-PubChem}, Morgan fingerprints~\cite{Fingerprint-Morgan} and so on are typical types..

Molecular fingerprints~\cite{molecular-fingerprint-1, molecular-fingerprint-2} attempt to encode a molecule into a list of bits by the presence of certain chemical fragments from a pre-defined set of structural keys, which are simplification or abstraction of substructure patterns and need to be identified by domain knowledge. The length of bits relies on the number of chemical fragments contained in the target molecule, and each index in molecular fingerprints denotes the key of chemical structure. Molecular fingerprints are most useful when components of each molecule are likely to be covered in the structural keys set, however, face challenges when molecules contain substructures out of the keys set.

To get rid of the full understanding of the keys set, the most straightforward yet simplest strategy is to traverse all possible (sub)structures by random walk model and then map each known or unknown structure into a unique code based on embedding techniques. In this way, any substructure is only determined by its contained atoms as well as the linked mode in each atom chain, regardless of its judicious chemical definition or function from the prior knowledge. Here, the worthy issue notable for each mined substructure lies in the topological \emph{\textbf{scale}}, which corresponds to the length $r$ of each atom chain via random walk. Particularly, we denote the scale of the most expressive substructures as the \emph{\textbf{typical scale}} in the targeted molecule.

\textbf{Present Work.} In this paper, we decompose a molecular graph into a series of $r$-scale anonymous atom chains and formulate a new molecular fingerprint method \emph{\textbf{Anonymous Pattern Molecular Fingerprint}} (abbreviated as \emph{\textbf{Anonymous-FP}}) based on such structures, where $r$ denotes the length of an atom chain from the source atom to the ending atom. Our methodology consists of two steps: sampling anonymous atom chains by random walks and coding anonymous atom chains by embedding techniques.

\textbf{Step 1. Sampling anonymous atom chains.} We sample $t$ times $r$-scale random walks beginning at each atom in the molecule and collect them as a set, which represents the atom chain decomposition. It is obvious that the probability of an $r$-scale atom chain occurring decreases with the increasing distance $r$. To avoid distribution sparsity, we take a special encoding mode named anonymous-based random walk~\cite{Anonymous-walk} to transfer each atom chain into its anonymous pattern (i.e., anonymous atom chain) as a sequence of indexes. Each position of such sequence denotes the order of the first occurrence of the corresponding atom in the chain. This schema makes \emph{Anonymous-FP} suitable for molecules that are absent of global structural keys and even labels of some atoms. Moreover, this schema is also computational complexity efficient because anonymous coding is usually statistically significant, especially in understanding long $r$-scale atom chains with sparse distribution.

\textbf{Step 2. Encoding anonymous atom chains.} To qualify similarity between two molecules with different anonymous atom chain decomposition faces the challenge of different chain amounts. Inspired by a Natural Language Processing (\emph{\textbf{NLP}}) document embedding technique \emph{\textbf{PV-DBOW}}~\cite{PV-DBOW}, which encodes each document into a vector representation, we treat a molecule as a document and anonymous atom chains as interacted words inside. Then we embed each molecule into Euclidean space as a vector and denote such fixed-size vector as a molecular fingerprint named \emph{Anonymous-FP}. Our institution origins from the Similarity Property Principle (\emph{\textbf{SPP}})~\cite{similarity-property-principle, similarity-property-principle=2}, which points out that molecules express similar physicochemical properties if they share similar structural features, similar anonymous atom chain decomposition as well as proximity in embedding space.

We evaluate the efficiency of \emph{Anonymous-FP} on property identification, such as property classification, using a series of real-world molecular datasets (MUTAG, PTC, PROTEINS, DD, and NCIs). We compare the performance of \emph{Anonymous-FP} with kernel methods (Graphlet kernel, Weisfeiler-Lehman kernel), embedding methods (Graph2vec, AWE), and Graph Neural Networks (PATCHY-SAN, GraphSAGE). The experiment results indicate that our \emph{Anonymous-FP} shows a considerable advantage over others in terms of classification accuracy. We present a systematic analysis of the correlation between molecular graph representation power and the atom chain length $r$ as well as sampling number $t$. Meanwhile, the scale that induces the highest accuracy is followed as the typical scale. A more interesting discovery is then proposed: in all NCI datasets, the classification accuracy will achieve the best and go over $93 \%$ when $r$ turns into $8$.

\section{Related Works}\label{Section-Related-Work}

\begin{figure*}[t]\label{Figure-Fingerprint}
\centering
\includegraphics[height=6.5cm,width=19cm]{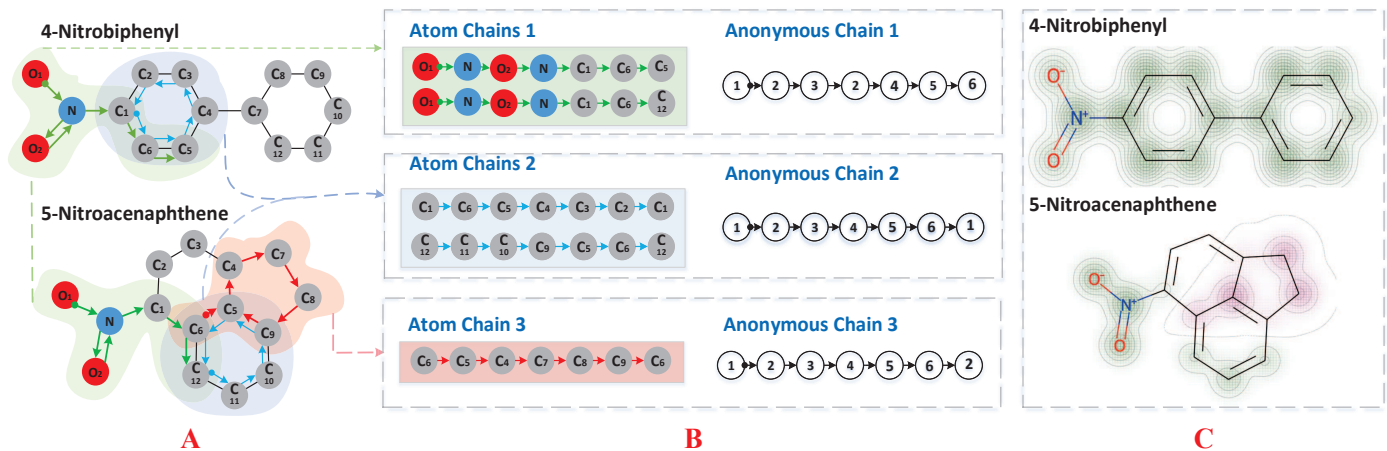}
\caption{
An overview of the relation between \emph{atom chain} and \emph{anonymous atom chain}. As shown in subfigure (\textcolor{red}{A}), oxygen atoms, nitrogen atoms, and carbon atoms are colored red, blue, and grey.  Each of them is arrayed differently from 1 to 12, respectively. In subfigure (\textcolor{red}{B}), anonymous atom chains 1-3 correspond to 5 different 6-scale atom chains, which are sampled randomly from these two molecules and covered with light green, light blue, and light red shadows. Among these three anonymous atom  chains, anonymous atom chain 1 and anonymous atom chain 2 are together shared by both molecules, while anonymous atom chain 3 is unique to 5-Nitroacenaphthene. This also implies the difference between these two molecules and could be roughly shown in subfigure (\textcolor{red}{C}).}
\end{figure*}

In this section, we review the related works for the primary methods used in this paper, including molecular fingerprints, basic definitions of an unweighted graph, molecular graph embedding, and \emph{PV-DBOW} technique in \emph{NLP}.

\textbf{Molecular Fingerprints.}
Molecular fingerprints are essential cheminformatics tools dedicated to searching, describing, and validating molecular structural characteristics through vectorial representation and comparison. Diverse pioneering fingerprints could be roughly classified into four categories, including substructure keys fingerprints, topological fingerprints, pharmacophore fingerprints, and other types.

Substructure keys fingerprints encode each molecule into a bit string based on the presence of substructures from a set of structural keys but lose effectiveness when absenting substructures from the keys set. MACCS~\cite{Fingerprint-MACCS}, PubMed fingerprints~\cite{Fingerprint-PubChem}, and BCI fingerprints~\cite{Fingerprint-BCI} fingerprints are typical ones. Topological fingerprints (such as Molprint2D~\cite{Fingerprint-Molprint2D}, ECFP~\cite{Fingerprint-ECFP}, and MP-MFP~\cite{Fingerprint-MP-MFP}) look for atom chains and then hashing everyone of them to create fingerprints. Pharmacophore fingerprints take account of molecular features from a list of targeted features~\cite{Fingerprint-Pharmacophore}. Other types usually generate fingerprints employing the canonical SMILES~\cite{Fingerprint-LINGO}, protein-ligand interactions~\cite{Fingerprint-Protein-Ligand-interactions}, and other structural interactions.

\textbf{Graph Representation.}
Let $G = (\mathcal{V}, \mathcal{E})$ denote an unweighted graph with $n$ vertices in set $\mathcal{V}$ and $m$ edges in set $\mathcal{E} \subseteq \mathcal{V} \times \mathcal{V}$, the adjacency matrix $\textrm{A} \in \mathbb{R}^{n \times n}$ encodes the vertex-wise connection of the graph and is defined as follows:
\begin{equation}
     \textrm{A}_{i, j}=
     \left\{
     \begin{aligned}
      1                 &,  \ if \ (v_{i}, v_{j}) \in \mathcal{E} \\
      0                 &,  \ otherwise
      \end{aligned}.
      \right.
\end{equation}

And the degree $d_{v_{i}}$ of vertex $v_{i}$ is defined as the sum of entries in $i$-th row from adjacency matrix $\textrm{A}$, which is exactly the number of 1-hop neighbors for $v_{i}$:
\begin{equation}
     d_{v_{i}} = \sum\limits_{j=1}^{n} \textrm{A}_{i, j}.
\end{equation}

The transition matrix $\textrm{P}$ records all the transition probabilities $Prob (v_{j} | v_{i})$ for an agent moving from vertex $v_{i}$ to anyone in its 1-hop neighborhood, and each entry $\textrm{P}_{i, j} = Prob (v_{j} | v_{i})$ satisfying
\begin{equation}
     \textrm{P}_{i, j} =
     \left\{
     \begin{aligned}
      \frac{1}{d_{v_{i}}}   &,  \ if \ (v_{i}, v_{j}) \in \mathcal{E} \\
      0                     &,  \ otherwise
      \end{aligned}.
      \right.
\end{equation}
Clearly, for each vertex $v_{i}$,
\begin{equation}
     \sum\limits_{j=1}^{n} \textrm{P}_{i, j} = 1.
\end{equation}

\textbf{Molecular Graph Embedding.} Molecular graph embedding is related to vector representation for molecules. It maps molecular graph $G$ into a $d$-dimension vector in Euclidean space, i.e.,
\begin{equation}
  \varphi : G \rightarrow \mathbb{R}^{d},
\end{equation}
where $\varphi$ denotes an embedding function.

\textbf{PV-DBOW.} In Natural Language Processing (\emph{\textbf{NLP}}), \emph{PV-DBOW} technique is used for unsupervised embedding sentiment in the level of the document. More specifically, given a document set $\mathcal{D} = \{D_{1},\ldots,D_{N}\}$ with a set of words $\mathcal{V}=\{\omega_{1}, \ldots, \omega_{\nu}\}$, for the target document $D_{i}\in \mathcal{D}$ which contains a sequence of words $\mathcal{V}_{i} = \{\omega_{1},\ldots,\omega_{l}\} \subseteq \mathcal{V}$, the goal is to learn low-dimension vector representation $\boldsymbol{D}_{i}$ for document $D_{i}$ by maximizing the following log probability:

\begin{equation}\label{pv-dbow}
  \sum\limits_{\omega_{j} \in \mathcal{V}_{i}}\log\bold{Pr}(\omega_{j}|D_{i}).
\end{equation}

The conditional probability $\bold{Pr}(w_{j}|D_{i})$ above is defined as softmax function:
\begin{equation}\label{softmax}
  \bold{Pr}(\omega_{j}|D_{i})=
  \frac{\exp(\boldsymbol{D}_{i}\cdot\boldsymbol{\omega}_{j})}
  {\sum\limits_{\omega_{m} \in \mathcal{V}}\exp(\boldsymbol{D}_{i}\cdot\boldsymbol{\omega}_{m})},
\end{equation}
where $\boldsymbol{\omega}_{j}$ is the corresponding representation vector of $\omega_{j}$.
% and $|\mathcal{V}|$ is the number of all words across all documents in $\mathcal{D}$.

\section{Methodology}
In this section, we introduce the details of the new proposed molecular fingerprint method \emph{Anonymous-FP}. In this method, each molecule from molecules set $\mathcal{G} = \{G_{1}, \ldots,G_{N}\}$ is decomposed into a set of $r$-scale \emph{\textbf{atom chain}}s, then we transform them into $r$-scale \emph{\textbf{anonymous atom chain}}s and embed molecule graph into high dimensional vector space.

\subsection{Atom Chain and Anonymous Atom Chain}
In a molecular graph $G_{i} = (\mathcal{V}_{i}, \mathcal{E}_{i})$ with the adjacent matrix $\mathbf{A}$ and transition matrix $\mathbf{P}$, the $r$-scale \emph{atom chain} is denoted as a Markov chain $w = (v_{0}, v_{1}, \ldots, v_{r})$, which is derived by such a process that an agent walks from the root atom $v_{0}$ to the end $v_{r}$ step by step.

Then, the \emph{anonymous atom chain} transforms each atom chain into a sequence of integers recording positions that appear first. More specifically, The anonymous atom chain $a$ for $r$-scale atom chain $w$ is a sequence of integers defined by operator $\psi$,
\begin{equation}\label{anonymous_operate}
  a = \psi (w) = [f (v_{0}), f (v_{1}), \ldots, f (v_{r})],
\end{equation}
in which $f$ is the position function such that $f(v_{i}) = |(v_{0}, \ldots, v_{\hat{i}})|$, where $\hat{i}$ is the smallest integer such that $v_{\hat{i}} = v_{i}$.

Supposing that at each root vertex in $G_{i}$, agent samples $t$ $r$-scale atom chains and collects them into a set as the structural decomposition for molecular graph $G_{i}$, denoted by $\mathcal{W}_{i} = \{w_{i}^{1}, \ldots ,w_{i}^{t}\}$. This atom chain set corresponds to an anonymous atom chain set $\mathcal{A}_{i} = \{a_{i}^{1}, \ldots ,a_{i}^{{\tau}_{i}}\}$, and obviously $|\mathcal{A}_{i}| \leq |\mathcal{W}_{i}|$.

Now we collect each molecular structural decomposition $\mathcal{W}_{i}$ into a union, denoted by
\begin{equation}
  \mathcal{W}  = \bigcup \limits_{i=1}^{N} \mathcal{W}_{i} = \{w_{1}, \ldots ,w_{\mu}\}.
\end{equation}

Then accordingly, the union of anonymous atom chains set is denoted as
\begin{equation}
\mathcal{A} = \bigcup \limits_{i=1}^{N} \mathcal{A}_{i} = \{a_{1}, \ldots ,a_{\nu}\}.
\end{equation}

The process of transforming each atom chain into its anonymous pattern is shown schematically in Figure~\ref{Figure-Fingerprint}. The basic idea in pattern translation origins from two reasons.

(1) Enhance the representation of unknown structures. In various pioneer molecular fingerprint methods, there always requires a full understanding about atoms, groups, or other substructures before generating fingerprints. However, in an anonymous pattern, an observer that conducts random walks records each atom only by its first occurrence in a random walk, regardless of its real atom category. This may help transfer any well-known or less-known substructure in chemistry and bioinformatics into a unique numerical sequence under a consistent rule.

\begin{figure}[t]
\centering
\includegraphics[height=6.5cm,width=9cm]{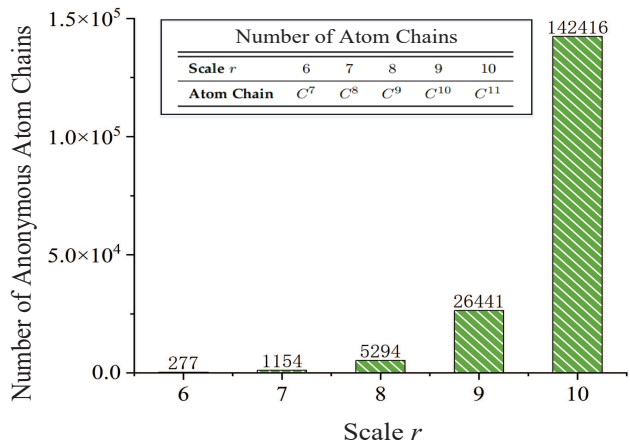}
\caption{Statistics of possible atom chains and anonymous atom chains when the targeted molecules are with $C$ classes atoms ($C > 10$).
}
\label{Figure-Number-AWS}
\end{figure}

(2) Reduce the computational complexity. For each $r$-scale atom chain $w = (v_{0}, v_{1}, \ldots, v_{r})$, the occurring probability is
\begin{equation}
P(w) = \prod \limits_{i=0}^{r-1} \mathrm{P}_{i, i+1},
\end{equation}
where the operational symbols follow the definitions in Section~\ref{Section-Related-Work}. Accordingly, the probability of choosing anonymous pattern $a = \psi (w)$ in $G_{i}$ equals
\begin{equation}
P(a) = \sum \limits_{w \in W_{i}, a = \psi (w)} P(w).
\end{equation}
In addition, we use the statistics in Figure~\ref{Figure-Number-AWS} to verify the simplification when conducting an anonymous pattern. Here we define the scale $r$ ranging from 6 to 10, and the overall atom categories $C > 10$. The number of possible atom chains and the number of anonymous atom chains is reported in table and histograms, respectively. With an increasing $r$, there faces an exponential rise in the number of possible atom chains, which equals $C^{r+1}$ and is greater than the number of $r$-scale anonymous atom chains. This result may be attributed to the fact a mass of atom chains with sparse distribution are all compressed into bits of anonymous atom chains, such that the overall computational complexity reduces significantly.

%\begin{figure}[t]
%\centering
%\includegraphics[height=6cm,width=9.5cm]{Figure-AW-Propotion.eps}
%\caption{The proportion of anonymous atom chains with different labeled nodes in all possible $r$-scale anonymous atom chains.
%}
%\label{Figure-Number-AWS}
%\end{figure}

\subsection{Anonymous-FP}
We propose a novel molecular fingerprints methodology on the basis of anonymous atom chains mined from molecules. In our work, a \emph{NLP} technique \emph{PV-DBOW} is adopted to encode each molecule as fixed-length vector embedding, and regard anonymous atom chains as words contained in a document (i.e., molecule). Our ideology comes from the Similarity Property Principle (\emph{SPP})~\cite{similarity-property-principle, similarity-property-principle=2} that molecules with similar anonymous atom chains share proximity in embedding space.

In mathematical framework, we suppose a $d$-dimension vector $\boldsymbol{G}_{i}$ as the representation for molecule $G_{i}$, and $\nu \times d$ matrix $\textbf{M}$ as encoding for anonymous atom chains set $\mathcal{A} = \{a_{1}, \ldots ,a_{\nu}\}$, where each row vector $\textbf{a}_{i}$ corresponds to anonymous atom chain $a_{i}$.

The global object is to embed a targeted molecule $G_{i}$ into $d$-dimension vector $\textbf{G}_{i}$ by minimizing the objective function
\begin{equation}\label{Target}
  \mathcal{L} = - \sum\limits_{a_{j}\in \mathcal{A}_{i}}\log{Prob}(a_{j}|G_{i}).
\end{equation}

The conditional probability ${Prob}(a_{j}|G_{i})$ above is defined as a softmax function:
\begin{equation}\label{softmax}
  {Prob}(a_{j}|G_{i}) =
  \frac{\exp(\boldsymbol{G}_{i} \cdot \boldsymbol{a}_{j})}
  {\sum\limits_{a_{m} \in \mathcal{A}} \exp(\boldsymbol{G}_{i} \cdot \boldsymbol{a}_{m})},
\end{equation}
where $\boldsymbol{G}_{i}$ and $\boldsymbol{a}_{m}$ are corresponding representation vectors of $G_{i}$ and $a_{m}$.
% and $\nu$ is the volume of anonymous chains set $\mathcal{A}$.

Since the volume $\nu$ for anonymous atom chains set $\mathcal{A}$ tends to be very large, the enumeration part of the softmax item (\ref{softmax}) requires a large amount of computing resources. Thus a negative sampling method~\cite{Negative-Sampling} is taken to approximate the log probability (\ref{Target}), which randomly samples a small portion of anonymous atom chains $\tilde{\mathcal{A}}_{i}^{a_{j}}$ as negative samples out of targeted molecule $\boldsymbol{G}_{i}$, i.e.,
\begin{equation}\label{negative-samples}
  \tilde{\mathcal{A}}_{i}^{a_{j}} = \mathcal{A}_{i} / \{a_{j}\}.
\end{equation}
Only target anonymous atom chain $a_{j}$ and negative samples are updated instead of all the elements from set $\mathcal{A}$ in the iterative training. This strategy would be efficient, especially for cases where tasks face huge computational complexity. Thus the objective function (\ref{Target}) could be rewritten as the following:
\begin{equation}\label{Target2}
\begin{aligned}
  \mathcal{L}   = & - \sum\limits_{a_{j}\in \mathcal{A}_{i}}\log{Prob}(a_{j}|G_{i}) \\
                = & \log \sigma (\boldsymbol{a}_{j} \cdot \boldsymbol{G}_{i}) + \sum\limits_{k=1}^{K} \mathbb{E}_{a_{k} \in \tilde{\mathcal{A}}_{i}^{a_{j}}}\log \sigma (-\boldsymbol{a}_{k} \cdot \boldsymbol{G}_{i}) \\
                = & \log \sigma (\boldsymbol{a}_{j} \cdot \boldsymbol{G}_{i}) + \sum\limits_{k=1}^{K} \mathbb{E}_{a_{k} \in \mathcal{A}_{i} / \{a_{j}\}}\log \sigma (-\boldsymbol{a}_{k} \cdot \boldsymbol{G}_{i})
\end{aligned}
\end{equation}
%- \sum\limits_{j=1}^{\nu} \{ \log \sigma (\boldsymbol{a}_{j} \cdot \boldsymbol{G}_{i}) + \log \sigma (- \boldsymbol{a}_{j} \cdot \boldsymbol{G}_{i})\\
%e                    & + \sum_{a_{k} \in \tilde{A_{j}}} \log \sigma (- \boldsymbol{a}_{k} \cdot \boldsymbol{G}_{i}) \},
where $\sigma (x) = \frac{1}{1 + \exp{(-x)}}$ denotes the sigmoid function, $\boldsymbol{a}_{k}$ is the vectorial embedding of negative sample $a_{k}$ sampled from $\tilde{\mathcal{A}}_{i}^{a_{j}}$ for $K$ times. We optimize this loss function (\ref{Target2}) with stochastic gradient descent and update $\boldsymbol{G}_{i}$ and $\boldsymbol{a}_{j}$. After the learning process finishes, we refer to this $d$-dimension vector $\boldsymbol{G}_{i}$ as fingerprint \emph{Anonymous-FP} for molecule $G_{i}$.

\emph{Anonymous-FP} has absorbed the advantages of substructure keys fingerprints, topological fingerprints, pharmacophore fingerprints, and other fingerprint types.
Here the reason is twofold. On the one hand, it restates and slightly extends substructure keys fingerprints and topological fingerprints as all substructures are sampled via random walk model as atom chains. Then atom chains are transferred as their anonymous pattern so that the reliance on prior knowledge about concrete substructure keys as well as the atoms is partly released. On the other hand, \emph{Anonymous-FP} undertakes \emph{PV-DBOW} to generate fingerprints with regard to the targeted graph as a document and the contained anonymous atom chains as words inside the document. Thus the mechanisms that underlie structural interactions are implied in vectorial representations and this acts as the original starting point of this molecular fingerprint method.

\begin{figure*}[t]
\centering
\includegraphics[height=7cm,width=18cm]{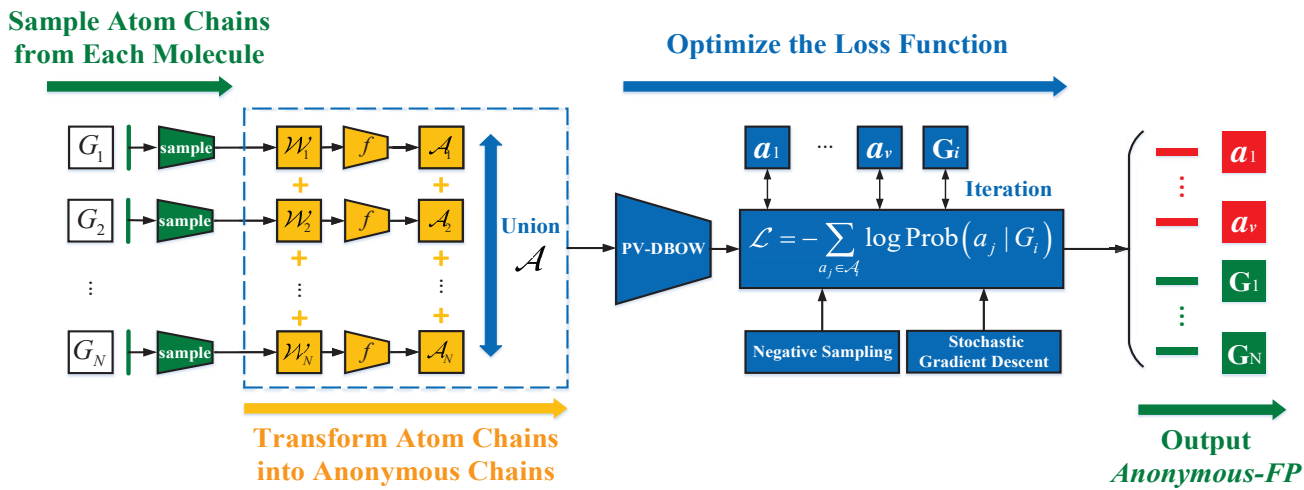}
\caption{The algorithm of \emph{Anonymous-FP}: Sampling atom chains from the initial graph data and then translating each atom chain into its anonymous atom chain. By taking the union of all the anonymous atom chain sets, the set $\mathcal{A}$ is built and then passes through the embedding model \emph{PV-DBOW}. After that, the \emph{Anonymous-FP} as outputs are provided iteratively.}
\label{Figure-Algorithm}
\end{figure*}

Algorithm~\ref{Anonymous-FP} and Figure~\ref{Figure-Algorithm} outline the framework of \emph{Anonymous-FP}. In its initialization, molecular embedding vector $\boldsymbol{G}_i$ as well as the anonymous atom chains vectors $\boldsymbol{a}_{j}$, $j = 1, \ldots, \nu$, are randomly preset by normal distribution $\mathcal{N}(0,0.01)$ first, then these embedding vectors are iteratively calculated by gradient descent until achieving convergence.

\begin{algorithm}[htbp]
	\renewcommand{\algorithmicrequire}{\textbf{Input:}}
	\renewcommand{\algorithmicensure}{\textbf{Output:}}
	\caption{\emph{\textbf{Anonymous-FP}}}
    \label{Anonymous-FP}
	
	\begin{algorithmic}[1]
		\REQUIRE molecules set $\mathcal{G}$ = $\{G_1, \ldots, G_N\}$; anonymous atom chains set $\mathcal{A} = \bigcup \limits_{i=1}^{N} \mathcal{A}_{i} = \{a_{1}, \ldots ,a_{\nu}\}$; scale $r$; vector dimension $d$
		\ENSURE \emph{Anonymous-FP} $\boldsymbol{G}_i$ for molecule $G_{i}$

        \STATE initialize $\boldsymbol{G}_{i} = [\epsilon_{s}]_{1 \times d}$, $\epsilon_{s} \sim \emph{N}(0,0.01)$
        \STATE initialize $\boldsymbol{a}_{j} = [\varepsilon_{s}]_{1 \times d}$, $\varepsilon_{s} \sim \emph{N}(0,0.01)$, $j = 1, \ldots, \nu$

        \FOR {each anonymous atom chain $a_{j}$ in $\mathcal{A}_{i}$}
            \STATE sample negative set $\tilde{\mathcal{A}}_{i}^{a_{j}}$ from set $\mathcal{A}$
            \STATE $\mathcal{L} =  \log \sigma (\boldsymbol{a}_{j} \cdot \boldsymbol{G}_{i}) + \sum\limits_{k=1}^{K} \mathbb{E}_{a_{k} \in \tilde{\mathcal{A}}_{i}^{a_{j}}}\log \sigma (-\boldsymbol{a}_{k} \cdot \boldsymbol{G}_{i})$

            \STATE $\boldsymbol{G}_{i}=\boldsymbol{G}_{i} -\alpha\frac{\partial \mathcal{L}}{\boldsymbol{G}_{i}}$
            % \STATE $\boldsymbol{a}_{k}=\boldsymbol{a}_{k} -\alpha\frac{\partial \mathcal{L}}{\boldsymbol{a}_{k}}$
            \STATE $\boldsymbol{a}_{j}=\boldsymbol{a}_{j} -\alpha\frac{\partial \mathcal{L}}{\boldsymbol{a}_{j}}$

        \ENDFOR
        \STATE \textbf{return} \emph{Anonymous-FP} $\boldsymbol{G}_{i}$
	\end{algorithmic}
\end{algorithm}

% \subsection{The Scale of Anonymous atom Chain}

\section{Experiments}\label{Section-Experiments}
In this section, to validate the efficiency of our proposed methodology, we conduct extensive experiments on molecular graph classification tasks for molecular graph datasets. This task is a supervised pattern with training data consisting of pairs of input data (i.e., \emph{Anonymous-FP} of each molecule) and desired output label (i.e., target physicochemical property). It shows our method could achieve superior performance compared with several well-used baselines. A brief discussion of the interaction between the expressiveness of \emph{Anonymous-FP} and the full exploration of walk-driven samples will also be provided in this following part.

\subsection{Datasets}
\emph{Anonymous-FP} is tested on a series of real-world molecule graph datasets: \textbf{NCI-1}, \textbf{NCI-109}, \textbf{PROTEINS}, \textbf{DD}, \textbf{MUTAG} and \textbf{PTC}. Each dataset belongs to a certain type of specific physicochemical property with the labels active and inactive. The statistics are covered in Table \ref{table-1-dataset1} and the brief descriptions are as follows.

\begin{itemize}
    \item \textbf{NCI-1}, \textbf{NCI-109}~\cite{dataset-nci} are datasets of chemical compounds divided by the anti-cancer property (active or negative). These datasets have been made publicly available by the National Cancer Institute (NCI).
    \item \textbf{PROTEINS}~\cite{dataset-protein} is a set of protein graphs where nodes represent secondary structure elements and edges indicate neighborhood in the amino-acid sequence or in 3-dimension space.
    \item \textbf{DD}~\cite{dataset-dd} is a dataset of protein structures where nodes represent amino acids and edges indicate spatial closeness, which is classiﬁed into enzymes or non-enzymes.
    \item \textbf{MUTAG}~\cite{dataset-mutag} is a dataset of aromatic and heteroaromatic nitro compounds labeled according to whether they have a mutagenic effect on bacteria or not.
    \item \textbf{PTC}~\cite{dataset-ptc-mr} consists of graph representations of chemical molecules labeled according to carcinogenicity for male and female rats.
\end{itemize}

\begin{table}
\centering
\caption{Statistics of the benchmark graph datasets. The columns are the name of the dataset, the number of positively labeled graphs, the number of graphs, the number of classes, and the average number of nodes/edges.}
\begin{tabular}{lp{0.8cm}p{0.5cm}p{0.5cm}ll}
\toprule
\toprule
\textbf{Dataset}    &\textbf{Positive}  & \textbf{Total} & \textbf{Class}   & \textbf{Ave. Node} & \textbf{Ave. Edge} \\
\midrule
\textbf{NCI-1}       &2055               & 4110          & 2                 & 29.87                 & 32.30\\
\textbf{NCI-109}     &2063               & 4126          & 2                 & 29.68                 & 32.13\\
\textbf{PROTEINS}   &556                & 1112          & 2                 & 39.06                 & 72.82\\
\textbf{DD}         &589                & 1178          & 2                 & 284.32                & 715.66\\
\textbf{MUTAG}      &94                 & 188           & 2                 & 17.93                 & 19.79\\
\textbf{PTC}        &172                & 344           & 2                 & 14.29                 & 14.69\\
\bottomrule
\end{tabular}
\label{table-1-dataset1}
\end{table}

\subsection{Baselines}\label{sec:Baselines}
To fully illustrate the notable performance of our model, we compare it with a series of baselines.

\begin{itemize}
  \item \textbf{Graphlet kernel}~\cite{GK-graphlet-kernel}: Graphlet kernel (GK) measures graph similarity by counting common $k$-node graphlets, and this ensures the computation complexity restricted in ploynomial time.

  \item \textbf{Weisfeiler-Lehman kernel}~\cite{WL-graph_kernel}: Weisfeiler-Lehman kernel (WL) maps graph data into a Weisfeiler-Lehman sequence, whose node attributes represent graph topology and label information.  WL kernel is wildly used in isomorphism tests on graphs since the runtime scales linearly in the number of edges of the graphs and the length of the Weisfeiler-Lehman graph sequence.

  \item \textbf{Graph2vec}~\cite{graph2vec}: Graph2vec treats rooted subgraphs as words and graphs as sentences or documents, then it uses Skip-gram in \emph{NLP} to get explicit graph embeddings.

  \item \textbf{AWE}~\cite{AWE}: AWE uses anonymous random walks to embed entire graphs in an unsupervised manner, but it takes a different embedding strategy compared with our methodology. AWE leverages the neighborhoods of anonymous walks while our work focuses on the co-occurring anonymous walks in the global scale.

  \item \textbf{PATCHY-SAN}~\cite{PSCN}: Analogous to convolutional neural networks (CNNs)~\cite{GCN-kipf, GCN}, PATCHY-SAN (PSCN) proposes a framework to perform convolutional operations for arbitrary graph data.

  \item \textbf{GraphSAGE}~\cite{GraphSAGE}: GraphSAGE takes advantage of an inductive framework to calculate graph embeddings by sampling and aggregating 1-hop and 2-hop neighborhood features.

  % \item \textbf{GIN-0:} Graph Isomorphism Network (GIN-0)~\cite{GIN} focuses on building an efficient readout function and has obtained great achievements. It is shown that GIN-0 could achieve as powerful performance as the Weisfeiler-Lehman graph isomorphism test.

  %\item \textbf{GAT:} GAT~\cite{GAT} adds the attention mechanism to GCNs, which assigns different weights to different neighbor nodes. By weighted aggregation of first-order neighborhood features, it makes great performance in graph representation learning tasks.
\end{itemize}

\subsection{Implementation and Hyper-parameters}
In this paper, we use Python 3.6.12, Tensorflow 1.2.0, Scikit-learn 0.24.1, Numpy 1.22.1, and Networkx 2.6.3 as the computing environment and all experiments are conducted on the workstation with 2 INTEL XEON CPUs and 4 NVIDIA GeForce GTX1080Ti GPUs. We first randomly divide each dataset into 10 equal parts and choose 9 samples for training and 1 sample for testing the efficiency. For fair evaluation, we take the same size of \emph{Anonymous-FP} as 128 for all datasets. In fact, there is a tightly inherent association between \emph{Anonymous-FP} representation and hyper-parameters scale $r$ as well as sampling number $t$. To explore this kind of association, we regard \emph{Anonymous-FP} as a function of $r$ which ranges from short scale 6 to 10 incrementally, and of $t$ which arises from 10 to 160. Then we build a molecular graph classifier using Support Vector Machine with RBF kernel~\cite{SVM-Rbf} to test and verify the discriminative power of \emph{Anonymous-FP} and discuss the trend of the classification accuracy under the control of hyper-parameters $r$, $t$.

\subsection{Performance evaluation metrics}

% The output of a binary classification model can be primarily represented by four terms: (1) true positive (TP) defined as the number of true positively labeled chemicals that are correctly predicted as active by the model; (2) false positive (FP) as the number of true negatively labeled chemicals incorrectly predicted as inactive; (3) true negative (TN) as the number of true negatively labeled chemicals correctly predicted as inactive; and (4) false negative (FN) as the number of true positively labeled chemicals incorrectly predicted as inactive.

Most evaluation metrics are derived from these five terms: accuracy, precision, recall, F1-Score, and ROC-AUC.
\begin{itemize}
    \item \textbf{Accuracy}~\cite{evaluation-metrics-1, evaluation-metrics-2} calculates the probability of a model to make a correct prediction for the active or inactive items.
    \item \textbf{Precision}~\cite{evaluation-metrics-1, evaluation-metrics-2} estimates the probability of a model to make a correct active class prediction.
    \item \textbf{Recall}~\cite{evaluation-metrics-1, evaluation-metrics-2}, referred to the true positive rate or sensitivity, represents the fraction of correctly predicted active chemicals.
    \item \textbf{F1-Score}~\cite{evaluation-metrics-1, evaluation-metrics-2}, referred to as a balance of the Precision and Recall, ranges from 0 to 1. A higher F1-Score indicates a better classifier.
    \item \textbf{ROC-AUC}~\cite{ROC-AUC}. Receiver Operator Characteristic (ROC) curves are used to show how a predictor compares with the true outcome. Typically, the ROC curve reflects how sensitivity (true positive rate) changes with varying specificity (true negative rate) for various thresholds. The predictive capabilities of a variable are commonly summarized by the Area Under Curve (AUC), which can derived by the integral measure under the line segments.
\end{itemize}

%\begin{equation}\label{Formula-Accuracy}
%  Accuracy = \frac{TP + TN}{TP + FN + FP + TN}
%\end{equation}

%\begin{equation}\label{Formula-Precission}
%  Precision = \frac{TP}{TP + FP}
%\end{equation}

%\begin{equation}\label{Formula-Recall}
%  Recall = \frac{TP}{TP + FN}
%\end{equation}

%\begin{equation}\label{Formula-F1-Score}
%  F1-Score = \frac{2 Precision \times Recall}{Precision + Recall}
%\end{equation}

\subsection{Overall Results}

\begin{table*}[h]
\renewcommand\arraystretch{1.2}
\caption{Average classification accuracy (mean ± std $\%$) of our approach and baselines on real-world molecular datasets. The best result is marked in \textcolor{blue}{\textbf{bold}}.}
\centering
\begin{tabular}{|l|cccccc|}
\hline
\textbf{Algorithm}                                          &\textbf{NCI-1}     &\textbf{NCI-109}   &\textbf{PROTEINS} &\textbf{DD}       &\textbf{MUATG}    &\textbf{PTC} \\
\hline
\textbf{Graphlet kernel}~\cite{GK-graphlet-kernel}          &62.28 $\pm$ 0.29   &62.60 $\pm$ 0.19   &71.67 $\pm$ 0.55  &78.45 $\pm$ 0.26  &80.63 $\pm$ 3.07  &57.26 $\pm$ 1.41\\
\textbf{Weisfeiler-Lehman kernel}~\cite{WL-graph_kernel}    &80.13 $\pm$ 0.50   &80.22 $\pm$ 0.3    &72.92 $\pm$ 0.56  &77.95 $\pm$ 0.70  &81.66 $\pm$ 2.11  &56.97 $\pm$ 2.01\\
\textbf{Graph2vec}~\cite{graph2vec}                         &73.22 $\pm$ 1.81   &74.26 $\pm$ 1.47   &73.30 $\pm$ 2.05  &58.64 $\pm$ 0.01  &83.15 $\pm$ 9.25  &60.17 $\pm$ 6.86\\
\textbf{AWE}~\cite{AWE}                                     &62.72 $\pm$ 1.67   &63.21 $\pm$ 1.42   &70.01 $\pm$ 2.52  &71.51 $\pm$ 4.02  &	87.87 $\pm$ 9.76  &	59.14 $\pm$ 1.83\\
\textbf{PATCHY-SAN}~\cite{PSCN}                             &78.59 $\pm$ 1.89   & -                 &\textcolor{blue}{75.89} $\pm$ 2.76  &77.12 $\pm$ 2.41  &	\textcolor{blue}{92.63 $\pm$ 4.21} &60.00 $\pm$ 4.82\\
\textbf{GraphSAGE}~\cite{GraphSAGE}                         &74.73 $\pm$ 1.34   &74.17 $\pm$ 2.89   &74.01 $\pm$ 4.27  &75.78 $\pm$ 3.91  &	78.75 $\pm$ 1.18  &	-\\
% \textbf{GIN-0}~\cite{GIN}                                   &82.70 $\pm$ 1.70   &	-   &76.20 $\pm$ 2.80  & -   &	89.40 $\pm$ 5.60  & \textcolor{blue}{64.60 $\pm$ 7.00}\\
\hline
\textbf{\emph{Anonymous-FP}}                                &\textcolor{blue}{95.74 $\pm$ 0.96}     &\textcolor{blue}{95.95 $\pm$ 0.74}     &72.32	$\pm$ 4.77 &\textcolor{blue}{94.83 $\pm$ 1.67}                               &81.58 $\pm$ 4.85   &\textcolor{blue}{61.14 $\pm$ 5.29}\\
\hline
\textbf{Typical Scale $r$}                                  &8                  &8                  &9                  &8                  &10                  &9 \\
\textbf{Sampling Number $t$}                                    &40                 &70                 &40                 &110                &30                 &130  \\
\hline
\end{tabular}
\label{Table-Results}
\end{table*}

\begin{figure*}[h]
\centering
\includegraphics[height=9.8cm,width=15cm]{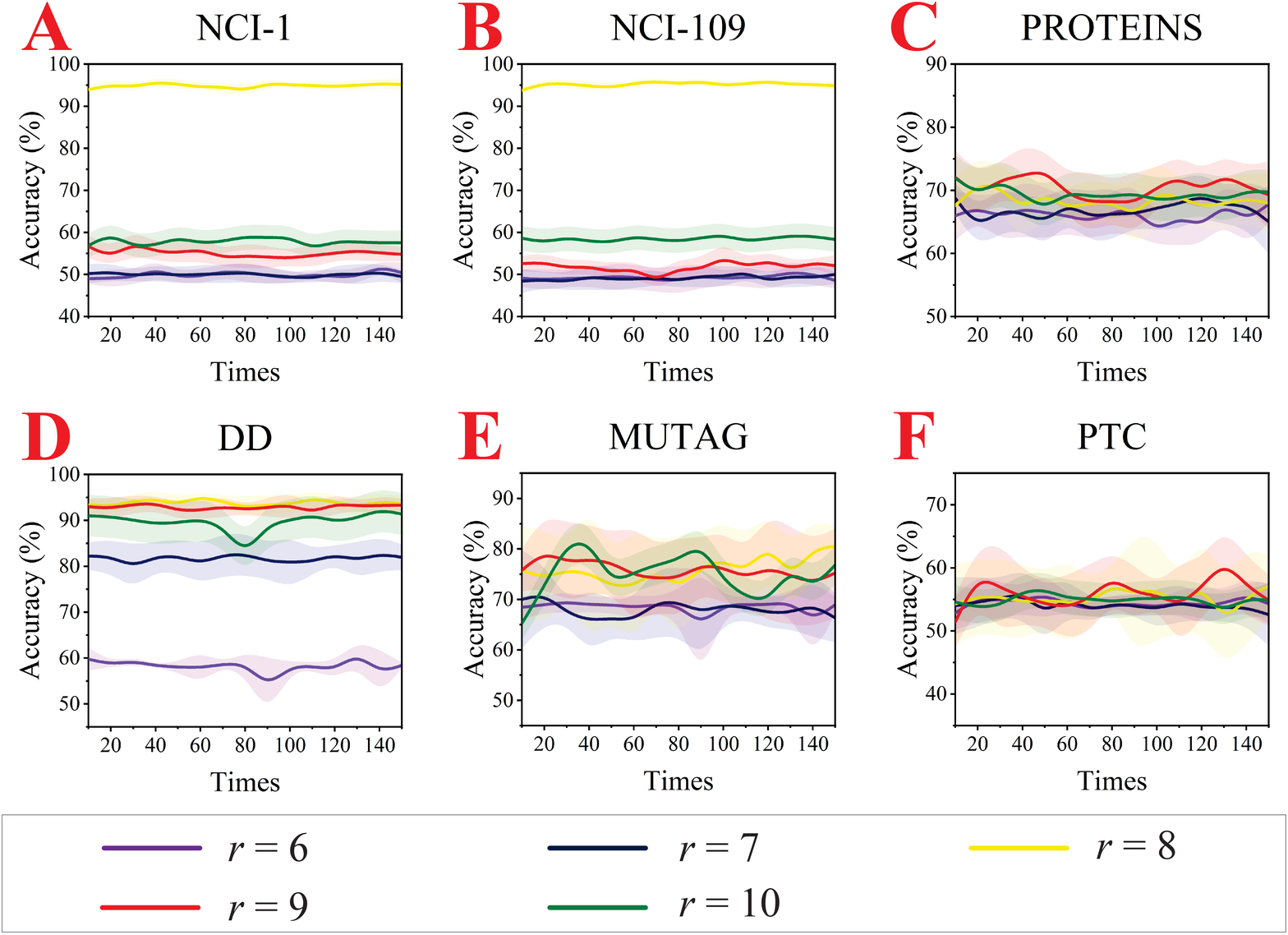}
\caption{Supervised graph classification performance of Support Vector Machine with RBF kernel classifiers for NCI-1 (\textcolor{red}{A}), NCI-109 (\textcolor{red}{B}), PROTEINS (\textcolor{red}{C}), DD (\textcolor{red}{D}), MUTAG (\textcolor{red}{E}), and PTC (\textcolor{red}{F}). Here the results are plotted as a function of the scale $r$ and the sampling number $t$. The shadow indicts the standard deviation of classification at each sampling point.
}
\label{Figure-Results}
\end{figure*}

\begin{figure*}[t]
\centering
\includegraphics[height=5cm,width=16cm]{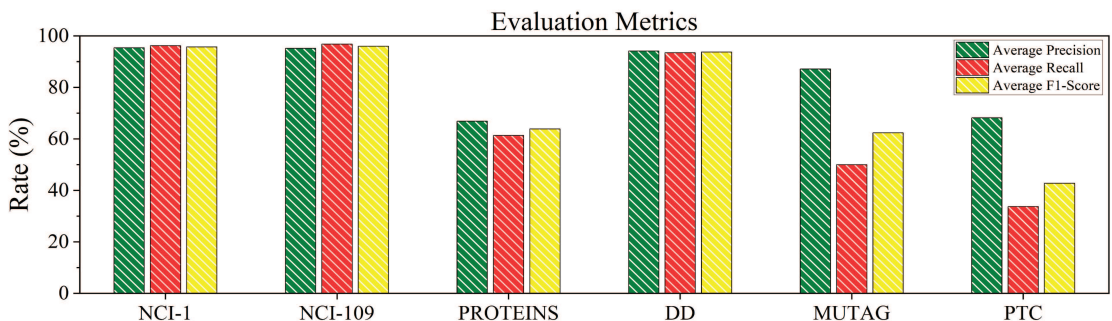}
\caption{The evaluation metrics with average precision, average recall, and average F1-score on different datasets are reported.}
\label{Figure-Evaluation_metrics}
\end{figure*}

\begin{figure}[t]
\centering
\includegraphics[height=6cm,width=6cm]{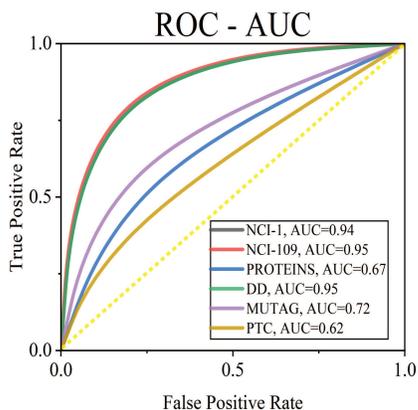}
\caption{The ROC curves and AUC values of different datasets. We present standard ROC curves of various datasets with different colors. Here the false positive rate is on the horizontal axis and the true positive rate is on the vertical axis. The diagonal dotted line denotes the identity.
}
\label{Figure-AUC}
\end{figure}

\subsubsection{Accuracy}

Table~\ref{Table-Results} summarizes the classification results calculated by baselines and \emph{Anonymous-FP}, meanwhile, the typical scale $r$ and sampling times $t$ that lead to the best performance of \emph{Anonymous-FP} are also provided in this table. From the table, we see that \emph{Anonymous-FP} performs powerful discriminative ability on NCI-1, NCI-109, DD, and PTC, with each classification accuracy far outweighing the best baseline in each dataset and equaling $95.74\%$, $95.95\%$, $94.83\%$, and $61.14\%$, respectively. However, \emph{Anonymous-FP} fails to outperform the baselines on PROTEINS and MUTAG datasets, with about $3.57\%$ and $10.95\%$ lower than the best result, respectively.

In addition, we proceed to the varying pattern of \emph{Anonymous-FP} classification performance corresponding to sequential $r$s and $t$s, and the trend is shown in Figure \ref{Figure-Results}. We first present classification performance as a function of scale $r\in [6, 7, 8, 9, 10]$, which depicts atom chains as well as anonymous atom chains in molecules from small scale to large scale.

When $r$ takes value as 6 or 7, the accuracy fluctuates around the initial value throughout the process of sampling number $t$ increasing for all datasets. In particular, the accuracy in NCI-1 or NCI-109 almost equals $50\%$, which means \emph{Anonymous-FP} exhibits little discriminative ability for these two balanced labeled datasets. For DD and PROTEINS, increasing sampling number $t$ still has no positive effects on classification accuracy when the scale $r =7$.

For middle-scale $r$ values, i.e., $r = 8$, it turns out that \emph{Anonymous-FP} shows a close relation to higher classification performance for all six datasets. As an overall view for NCI-1, NCI-109, and DD datasets when scale $r=8$, the classifications outperform the best accuracies. For NCI-1 and NCI-109, we could clearly see that the growth trend increases drastically by almost $30\%$ when scale $r$ turns into 8, with the curve reaching the maximum values $95.74\%$ and $95.95\%$ respectively and maintaining oscillating around the top. The best performances are also manifested when $r = 8$ for the DD dataset. While this performance does not hold for MUTAG, where the best is reached with the scale of $10$.

While for large $r$ scales, i.e., $9$ or $10$, \emph{Anonymous-FP} fails to show performance as competitive as that when scale $r$s equaling 8 for NCI-1, NCI-109. A notable phenomenon found is that the accuracy declines sharply once scale $r$ increases or decreases from $9$, and with the classification accuracies reaching values that are no more than $60\%$.

Therefore, due to the above analysis, it leads us to the fact that \emph{Anonymous-FP} is put into close relation with hyper-parameter anonymous atom chains scale $r$. In particular, \emph{Anonymous-FP} derived from 8-scale anonymous atom chains performs remarkable molecule discriminative ability. While this classification performance could disappear suddenly, especially for NCI-1 and NCI-109 when the scale receives a relatively small change, even $r$ increases from 8 to 9. But it is still necessary for further investigation to verify whether this phenomenon generally exists on other NCI molecule datasets.

\subsubsection{Precision, recall, F1-score, and ROC-AUC}

Precision, recall, F1-score, and ROC-AUC are also significant metrics to evaluate \emph{Anonymous-FP}. We show the results in Figure~\ref{Figure-Evaluation_metrics} and Figure~\ref{Figure-AUC}.

In Figure~\ref{Figure-Evaluation_metrics}, we compare the precision, recall, and F1-score of \emph{Anonymous-FP} on NCI-1, NCI-109, PROTEINS, DD, MUTAG, and PTC molecule datasets. On NCI-1, NCI-109, and DD, \emph{Anonymous-FP} can boost the precision, recall, and F1-score to more than 90\%. This indicts the present methodology can open horizons for the accuracy of predicted positive cases that are correctly real positives, real positive cases that are correctly predicted positive, and the balanced performance. On PROTEINS, MUTAG, and PTC, the precision is higher than the recall and F1-score in general. This means opting to model the molecules via \emph{Anonymous-FP} benefits the precision prediction of mass positive cases. The ROC-AUC in Figure~\ref{Figure-AUC} is used to determine the best model over a series of thresholds, where a model with a larger area under the curve (AUC) corresponds to better comprehensive performance. For NCI-1, NCI-109, and DD, the AUC is no less than 0.94, which is consistent with Figure~\ref{Figure-Evaluation_metrics} and Table~\ref{Table-Results}. The PROTEINS, MUTAG, and PTC reach AUC with 0.67, 0.72, and 0.62.

%\begin{table}[H]
%\centering
%\caption{The average evaluation metrics: average precision, average recall, and average F1-Score}
%{
%\begin{tabular}{|l|ccc|}
%\hline
%\textbf{Dataset}        & \textbf{Ave. Precision}   & \textbf{Ave. Recall}  & \textbf{Ave. F1-Score}\\
%\hline
%\textbf{NCI-1}          & 95.33                     & 96.21                 & 95.76                 \\
%\textbf{NCI-109}        & 95.16                     & 96.84                 & 95.98                 \\
%\textbf{PROTEINS}       & 66.89                     & 61.33                 & 63.87                 \\
%\textbf{DD}             & 94.15                     & 93.47                 & 93.76                 \\
%\textbf{MUATG}          & 87.16                     & 50.00                 & 62.38                 \\
%\textbf{PTC}            & 68.19                     & 33.75                 & 42.77                 \\
%\hline
%\end{tabular}}
%\label{Table-NCI-Series}
%\end{table}

\begin{figure*}[t]
\centering
\includegraphics[height=9.8cm,width=15cm]{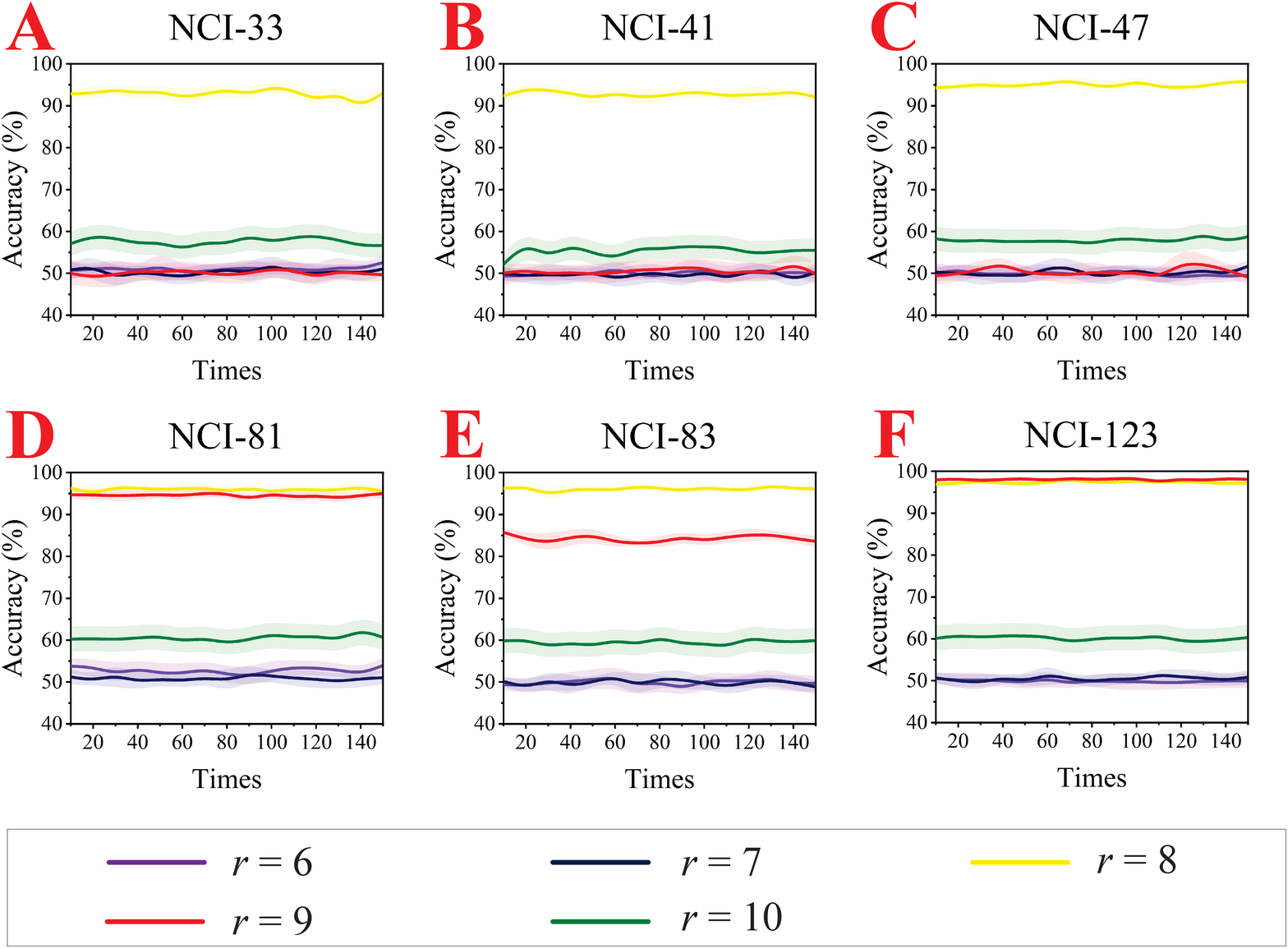}
\caption{
Supervised graph classification performance of Support Vector Machine with RBF kernel classifiers for NCI-33 (\textcolor{red}{A}), NCI-41 (\textcolor{red}{B}), NCI-47 (\textcolor{red}{C}), NCI-81 (\textcolor{red}{D}), NCI-83 (\textcolor{red}{E}), and NCI-123 (\textcolor{red}{F}).
}
\label{Figure-NCI}
\end{figure*}

\subsection{Additional Experiments on NCI Datasets}
In addition, we apply our proposed method to more NCI datasets, whose details are summarized in Table~\ref{Table-NCI-Series}. Each instance represents a set of molecules with active or inactive effects on particular cancer, and each set is separated by balanced prior label distribution. In these experiments, we pay attention to how the sampling number $t$ (arranging from 10 to 160), and anonymous atom chain scale $r$ (equalling 6, 7, 8, 9, or 10) affect the classification. Table~\ref{Table-NCI-results} shows the best result achieved by our proposed approach, the typical scale $r$, and the sampling time $t$ for each NCI dataset respectively.

\begin{figure*}[t]
\centering
\includegraphics[height=7cm,width=16cm]{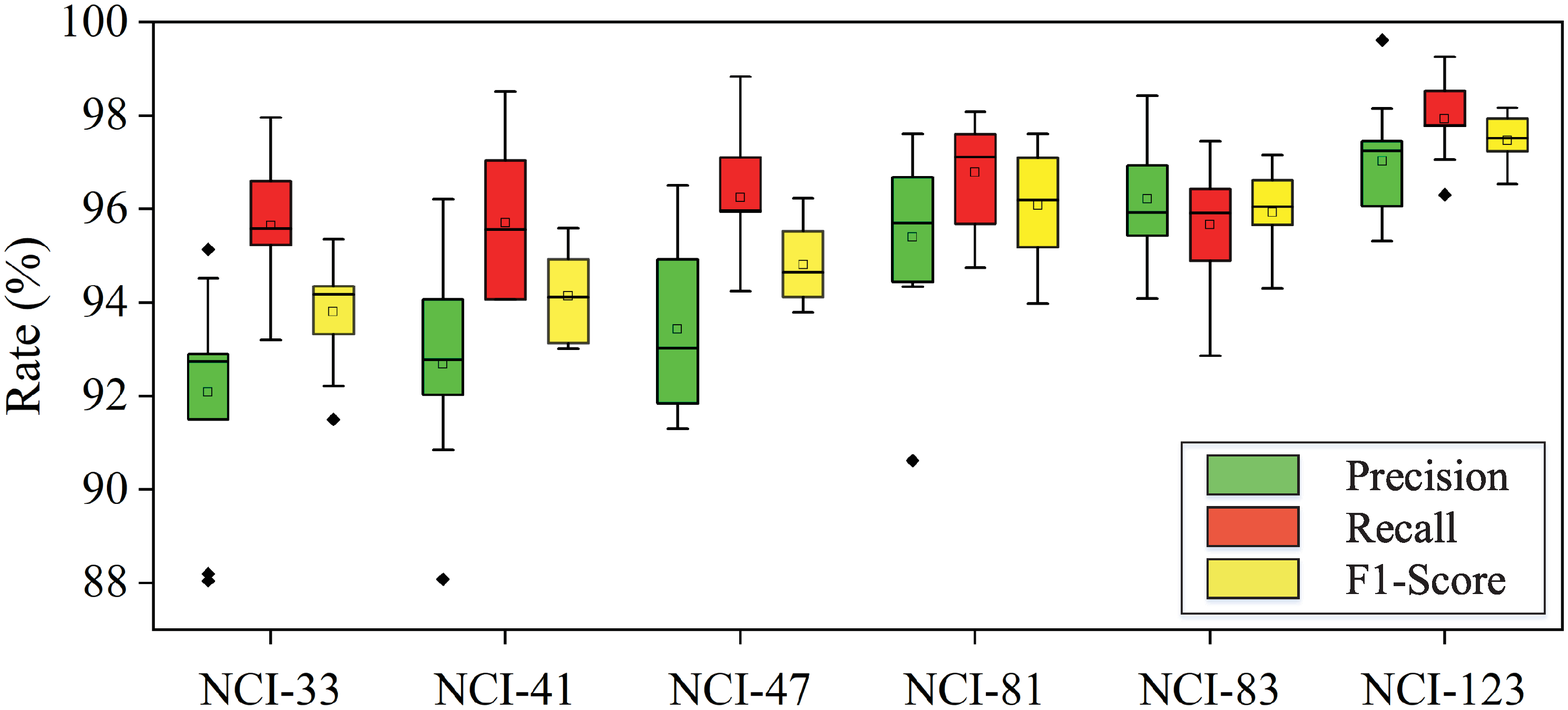}
\caption{The box-plots with precision, recall, and F1-score of 8-scale \emph{Anonymous-FP} on different datasets are reported. The box-plots consist of the most extreme values in the data set (maximum and minimum values), the lower and upper quartiles, the median, and the extreme outliers. In each box-plot, most values are gathered between the lower and upper whiskers except the outliers denoted as solid nodes. The lower and upper quartiles form the box, and the solid line represents the median. }
\label{Figure-Box}
\end{figure*}

In Figure~\ref{Figure-NCI}, for each dataset, it is clear that our method performs unsatisfactorily when $r = 7$ regardless of $t$'s values. This means sampling time $t$ has little effect on the results in this situation. Once scale $r$ turns to 8, the results of all NCI datasets arise sharply towards more than $93 \%$ and then are maintained around the best results as $t$ increases. Note that these 6 datasets react differently to scale $r = 9$: NCI-33, NCI-41, and NCI-47 suffer insensitive influence to scale $r = 9$, while the performance of NCI-81, NCI-83, or NCI-123 is able to level up when scale $r = 9$. In addition, the overall precision, recall, and F1-score, as complementary metrics, are also reported via bot-plots shown in Figure~\ref{Figure-Box}. In the end, the results from the above analysis explicitly support our inference: \emph{Anonymous-FP} derived from $8$-scale anonymous atom chains could better distinguish molecules, and this fact holds commonly for a mass of NCI databases.

% Tox21~\cite{dataset-Tox21, dataset-Tox21-2}: Tox21 is from a federal collaboration program ``Toxicology in the 21st Century (Tox21)''. Tox21aims to develop better toxicity assessment methods to quickly and efficiently test whether certain chemical compounds can potentially disrupt human body processes that may lead to negative health effects. Among these datasets, Tox21-AR and Tox21-AR-LBD are used to test toxicity for the androgen receptors, and Tox21-ER is to test toxicity for estrogen receptors.

\begin{table}[H]
\centering
\caption{Statistics of NCI datasets. The columns are the name of the dataset, the number of positively labeled graphs, the number of total graphs, and the description of the corresponding dataset.}
{
\begin{tabular}{lccl}
\toprule
\toprule
\textbf{Dataset}        & \textbf{Positive Graphs}  & \textbf{Total Graphs}   & \textbf{Tumor Description}    \\
\midrule
\textbf{NCI-33}          & 1467              & 2934              & Melanoma                      \\
\textbf{NCI-41}          & 1350              & 2700              & Prostate                      \\
\textbf{NCI-47}          & 1735              & 3470              & Central Nerv Sys              \\
\textbf{NCI-81}          & 2081              & 4162              & Colon                         \\
\textbf{NCI-83}          & 1959              & 3918              & Breast                        \\
\textbf{NCI-123 }        & 2715              & 5430              & Leukemia                      \\
%\textbf{Tox21-AR}       & 2715              & 5430              & Androgen Receptors                      \\
%\textbf{Tox21-AR-LBD}   & 2715              & 5430              & Androgen Receptors                      \\
%\textbf{Tox21-ER}       & 2715              & 5430              & Estrogen Receptors                      \\
\bottomrule
\end{tabular}}
\label{Table-NCI-Series}
\end{table}

\begin{table}[H]
\renewcommand\arraystretch{1.2}
\caption{Classification results (mean ± std $\%$), the typical scale $r$, and the sampling number $t$ for each NCI dataset.}
\centering
%\begin{tabular}{p{26 pt} p{25 pt} p{27 pt} p{27 pt} p{27pt} p{27 pt} p{31 pt}}
\begin{tabular}{|l|c c c|}
\hline
\textbf{Dataset}    &\textbf{Ave. Accuracy}  &\textbf{Typical Scale $r$}     &\textbf{Sampling Number $t$} \\
\hline
\textbf{NCI-33}     &\textbf{95.53 $\pm$ 1.17}  &\textbf{8}     &\textbf{100} \\
\textbf{NCI-41}     &\textbf{93.92 $\pm$ 1.30}  &\textbf{8}     &\textbf{20} \\
\textbf{NCI-47}     &\textbf{96.02 $\pm$ 0.92}  &\textbf{8}     &\textbf{70} \\
\textbf{NCI-81}     &\textbf{96.57 $\pm$ 0.85}  &\textbf{8}     &\textbf{140} \\
\textbf{NCI-83}     &\textbf{96.91 $\pm$ 0.57}  &\textbf{8}     &\textbf{130} \\
\textbf{NCI-123}    &\textbf{97.78 $\pm$ 0.63}  &\textbf{8}     &\textbf{100} \\
\hline
\end{tabular}
\label{Table-NCI-results}
\end{table}

\section{Conclusion and Further Outlook}

Molecules are usually in the form of vertex-edge topological graph structure, which is constructed by a collection of atom chains. The anonymous pattern of atom chains reflects strong associations between items within a molecule and carries the underlying semantics of the data. In this paper, we propose a novel molecular fingerprints method, \emph{Anonymous-FP}, for discriminating molecules using their embedded anonymous atom chains decompositions. The advantage of this approach lies in that, it leverages a \emph{NLP} technique \emph{PV-DBOW} to encode each molecule into a vector, which acts as a characterization of global molecule structures without the need of understanding each chemical symbol for each atom.

As a highlight, the scale $r$ of the anonymous atom chain plays an important role in the representation of molecular properties. Typically, the typical scale $r=8$ could significantly level up the discriminative accuracy for a series of datasets and this interesting phenomenon holds pretty generally for more NCI datasets.
However, the potential reason for scale $r=8$ commonly promoting representation is still unknown as of yet. Furthermore, it is also interesting to gain more insight into more effective fingerprint designs that are preferable for larger molecule representation.

\section*{Acknowledgements}
This work is supported by the National Natural Science Foundation of China (Grant Nos. 62276013, 62141605, 62050132), the Beijing Natural Science Foundation (Grant No. 1192012), and the Fundamental Research Funds for the Central Universities.

\bibliographystyle{unsrt}
\bibliography{MolecularFingerprintbibfile}
\end{document}